\documentclass[aip,pop,showpacs,reprint]{revtex4-1}
\usepackage{amsmath,amssymb}
\usepackage{graphicx}
\usepackage{color}

\def\x{\bf x}

\def\E{\bf E}

\begin{document}

\title{Orbital-motion-limited theory of dust charging and plasma response}

\author{Xian-Zhu~Tang}
\email{xtang@lanl.gov}
\author{Gian Luca Delzanno}
\email{delzanno@lanl.gov}

\affiliation{Theoretical Division, Los Alamos National Laboratory, Los
  Alamos, NM 87545}

\date{\today}

\begin{abstract}
  The foundational theory for dusty plasmas is the dust charging
  theory that provides the dust potential and charge arising from the
  dust interaction with a plasma.  The most widely used dust charging
  theory for negatively charged dust particles is the so-called
  orbital motion limited (OML) theory, which predicts the dust
  potential and heat collection accurately for a variety of
  applications, but was previously found to be incapable of evaluating
  the dust charge and plasma response in any situation.  Here we
  report a revised OML formulation that is able to predict the plasma
  response and hence the dust charge. Numerical solutions of the new
  OML model show that the widely-used Whipple approximation of dust
  charge-potential relationship agrees with OML theory in the limit of
  small dust radius compared with plasma Debye length, but incurs
  large (order-unity) deviation from the OML prediction when the dust size becomes
  comparable with or larger than plasma Debye length. This latter case
  is expected for the important application of dust particles in a
  tokamak plasma.
\end{abstract}

\maketitle

\section{Introduction}

Two years before Irving Langmuir coined the term ``plasma'' as in
plasma physics~\cite{langmuir-pnas-1928}, he and Mott-Smith laid down
a foundational theory on the charging of a spherical and cylindrical
probe in a laboratory plasma~\cite{mott-smith-langmuir-pr-1926}, which
was necessary for interpreting the measurement of what became later
known as Langmuir probes~\cite{hutchinson-book-2005}.  In the ensuing
decades, the charging and dynamics of solid particulates immersed in
plasmas built the foundation for a new discipline in plasma physics --
dusty
plasmas~\cite{shukla-book-2001,fortov-book-2010,morfill-rmp-2009}, in
which the collective behavior of a group of dust particles in a plasma
environment is studied.  The physics of both dust in a plasma and
dusty plasmas finds applications in
space~\cite{whipple-rpp-1981,verheest-book-2000},
astrophysics~\cite{draine-araa-2003,mendis94}, and
laboratory~\cite{bouchoule-book-1999,bacharis10,krash11}.  The
fundamental dust-plasma interaction, which is also essential to
understand the collective behavior in a dusty plasma, includes (1)
charging of the dust by absorption of plasma particles so it becomes
subject to the electric field (${\bf E}$) via the electrical force
${Q}_d{\E}$ (${Q}_d$ is the dust charge); (2) heating of the dust via
the collection of plasma electron and ion energy fluxes so the dust
particulate can melt, evaporate, or simply sublimate; (3) dragging of
the dust via a frictional force enhanced by the Coulomb interaction
with the flowing background plasma.

To understand the dust dynamics and its change of state, one must
resort to a dust charging theory.  The most-widely used of such is the
Orbital-Motion-Limited (OML) theory, which dates back to the work of
Mott-Smith and Langmuir~\cite{mott-smith-langmuir-pr-1926}.  In the
1960s, Al'pert et al~\cite{alpert-book-1965} and
Laframboise~\cite{laframboise-thesis-1966} completed the current
formulation.  The OML theory is known to predict accurately the dust
potential for a small~\cite{lampe-jpp-2001,kennedy03} and not so small dust
grain~\cite{delzanno04,willis-etal-psst-2010,delzanno-etal-ieee-2013},
despite its simplifying assumption on collisionless ion orbit that
misses the absorption radius effect away from the dust
surface~\cite{bohm-etal-book-1949,allen-ps-1992}.  Surprisingly, the
existing formulation can not be used for predicting the plasma
response, namely solving the plasma potential $\phi$, as shown in
Allen et al~\cite{allen-etal-jpp-2000}.  As the result, OML theory to
date can not predict the fundamental quantity of dust charge.  In
practice, one has been using an idealized dust charge-potential
relationship due to Whipple~\cite{whipple-rpp-1981}, who computed the
dust capacitance using the conventional Debye shielding potential that
does not take into account the constraint of angular momentum
conservation in setting the plasma density near the dust surface.

The purpose of this paper is to present a revised OML formulation
that, for the first time, is able to predict the plasma response and
hence the dust charge. The OML predictions will then be contrasted
with the Whipple approximation to elucidate the missing physics in
dust charging for a spherical dust of comparable size with the Debye
length, which is of special importance to dust transport/survivability
in tokamaks~\cite{tang-delzanno-jfe-2007,delzanno-tang-pop-2014}.
This new OML formulation is also important for electron-emitting dust
particulates as it provides the basis for extending the OML theory to
positively charged dust~\cite{delzanno-tang-submitted-2014}.

As an approximation of the Orbital Motion (OM) theory which follows
the collisionless particle orbit via conservation of energy and
angular momentum, the OML theory will inevitably introduce
discrepancies in the evaluation of dust current/heat collection, and
in the plasma ion density evaluation. The former will give rise to a
discrepancy in dust potential, while the latter in plasma potential
and hence dust charge, in comparison with those predicted by the OM
theory.  The usefulness of the OML formulation is due to its
simplicity, and the relatively high accuracy for a variety of
applications where the dust size is not large compared with the plasma
Debye length.  \textcolor{black}{It must be clarified that the
  fatal breakdown of the previous OML formulation in the form
  of an imaginary ion density when applied to calculate the plasma
  potential and dust potential~\cite{allen-etal-jpp-2000} is not due
  to the OML approximation of ignoring the ion absorption radius
  effect.  The root cause is the OML ion density formula originally
  given in Ref.~\cite{alpert-book-1965}, which we find will produce an
  imaginary density as long as Debye shielding of a dust particle is
  in place.  In the appendix, after we present the corrected OML
  theory and its application in calculating the plasma potential and
  dust charge, we give a detailed account of how the previous OML
  formulation would break even in the regime of small dust size
  compared to Debye length, a limit commonly accepted for OML
  approximation to be highly accurate.}

The rest of the paper is organized as follows. In
section~\ref{sec:oml}, we briefly recall the key components of the OML
theory and its inability to predict dust charge in its current
formulation.  A corrected formulation on the OML ion density is shown
in section~\ref{sec:corrected-oml}, which completes the OML
formulation for calculating both dust potential and dust change.  In
section~\ref{sec:oml-result}, we apply the new OML model to evaluate
the plasma response and the dust charge, and contrast the results with
the widely used Whipple approximation.  The findings are
summarized in section~\ref{sec:summary}.

\section{Background on OML theory\label{sec:oml}} 

There are three essential ideas underlying OML theory. First, the
collection of plasma electrons and ions by the dust is governed by
their collisionless orbit. These are subject to two conservation laws:
energy [$E=m_\alpha \left (v_r^2 + v_t^2\right)/2 + q_\alpha\phi(r)$]
conservation and angular momentum ($J=m v_t r$) conservation. Here $m$
($q$) and $\phi$ are the mass (charge) of the particles of species
$\alpha$ ($\alpha=e,\,i$ labels electrons and ions, respectively) and
the plasma potential, while $r$, $v_r$ and $v_t$ are the radial
distance, radial velocity and tangential velocity in a spherical
reference frame centered on the dust grain. The radial motion of the
plasma particles is governed by a one degree-of-freedom Hamiltonian
with effective potential $\Phi_{\rm eff},$
\begin{align}
H  = \frac{1}{2} m_\alpha v_r^2 + \Phi_{\rm eff}(r),\,\,\,
\Phi_{\rm eff} \equiv \frac{1}{2} \frac{J^2}{m_\alpha r^2} + q_\alpha\phi.
\end{align}
Second, whether the plasma particle reaches the dust is governed by
$\Phi_{\rm eff}.$ For a typical negatively charged dust, $\phi$ is
negative and monotonically increasing with $r,$ so that $\Phi_{\rm
  eff}$ is positive and monotonically decreasing with $r$ for
electrons.  As the result, only electrons with $E>\Phi_{\rm
  eff}(r=r_d)$ from far away can reach the dust of radius $r_d.$ The
final and third idea, which is a simplifying approximation, is what
delineates OML from the more complete but rarely used Orbital Motion
(OM) theory \cite{laframboise-thesis-1966,kennedy03}.  The OML theory
approximates the $\Phi_{\rm eff}$ for ions also as a monotonic
function of $r,$ which results in the simplification that ions with
$v_r^2 > - 2 \Phi_{\rm eff}(r_d)/m_i$ from far away can reach and
charge the dust. This is a remarkable simplification since one does
not need $\phi(r)$ to evaluate the ion and the electron current
collected by the dust, which are 
\begin{align}
  I_e & = - e 4\pi r_d^2 n_{e0} \sqrt{\frac{k_B T_e}{2\pi m_e}} \exp\left(\frac{e\phi_d}{k_B T_e}\right),\\
  I_i & = Z e 4\pi r_d^2 Z n_{i0} \sqrt{\frac{k_B T_i}{2\pi m_i}}
  \left(1-\frac{Ze\phi_d}{k_B T_i}\right).
\end{align}
Here $e$ is the elementary charge, $k_B$
is the Boltzmann constant, $n_{e0}$ ($n_{i0}$) is the electron (ion)
density away from the dust, $n_{e0}=Zn_{i0}$, $T_{e,\,(i)}$ is the
electron (ion) temperature, $q_e=-e$, $q_i=Ze$, and
$\phi_d=\phi(r_d).$
Setting $I_i+I_e=0$, one can solve for the dust potential as a
function of $T_e/T_i,$ ion charge state $Z,$ and $m_e/m_i.$ For an
electron-proton plasma with $T_e=T_i,$ OML predicts $\phi_d = -2.5
(k_B T_e/e),$ which is in remarkable agreement with particle-in-cell
simulations~\cite{delzanno-etal-ieee-2013} that do not make the OML
assumption of ion $\Phi_{\rm eff}$ being monotonic.

\begin{figure}
\begin{centering}
\includegraphics[scale=0.3]{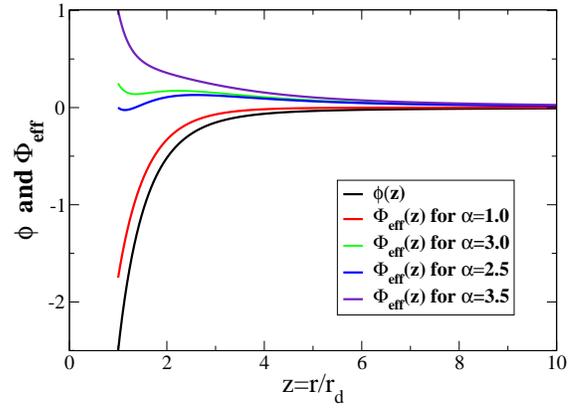}
\par\end{centering}
\caption{The effective potential $\Phi_{\rm eff}$ is shown for ions in
a hydrogen plasma
  with different values of $J$ or  $\alpha\equiv J^2/(2m_ik_B T_e).$ $\Phi_{\rm
    eff}$ can have extrema at $r_m>r_d$ that reflects some ions
  ($\alpha=2.5,3.0$). The OML approximation ignores this absorption
  radius effect, which can over-estimate the ion current ($\alpha=2.5$) and
the ion density between dust surface and the absorption radius ($\alpha=2.5, 3.0$).}
\label{fig:om-potential}
\end{figure}

The OML approximation of a monotonic ion $\Phi_{\rm eff}$ is known to
be violated for ions with a certain range of $J.$ This comes
about~\cite{allen-ps-1992} because the centrifugal potential energy
$J^2/(2m_i r^2)$ decreases with $r$ at precisely $1/r^2,$ but the
electrical potential energy $Ze\phi(r)$ increases at a faster rate
(exponential in $r$) in the Debye shielding region, before it
eventually asymptotes to $1/r^2$ for large $r.$ This can produce one
or multiple extrema in $\Phi_{\rm eff}$ away from the dust surface,
Fig.~\ref{fig:om-potential}, which can turn back some ions in a range
of $J$ at $r_m>r_d.$ In the literature this is known as the absorption
radius (at $r_m>r_d$) effect for certain ions. By neglecting this
subtlety, OML approximation would over-estimate the ion current to the
dust.  Interestingly, for dust size small and even comparable to Debye
length, this correction appears to be small and OML prediction of dust
potential remains reliable.

A far more serious problem was identified by Allen, Annaratone, and de
Angelis in 2000~\cite{allen-etal-jpp-2000} that the OML theory can not
predict the plasma response to the presence of a dust particle, which
requires the solution of the OML Poisson equation for the plasma
potential
\begin{align}
\nabla^2\phi = - {\varepsilon_0}^{-1}\left({Ze n_i^{OML} - e
  n_e^{OML}}\right), \label{eq:oml-poisson}
\end{align}
where $\varepsilon_0$ is vacuum permittivity.  Allen et
al.~\cite{allen-etal-jpp-2000} used the well-known OML ion and
electron density given by Al'pert, Gurevich, and Pitaevskii in
1965~\cite{alpert-book-1965},
\begin{widetext}
\begin{align}
\frac{n_e(z)}{n_{e0}} = &
\frac{1}{2}\left\{
1 + {\rm Erf}\left(\sqrt{\varphi-\varphi_d}\right)
+ \sqrt{1-z^{-2}}\left[1 - {\rm Erf}\left(\sqrt{\frac{\varphi-\varphi_d}{1-z^{-2}}}\right)
\right]
\exp\left[\frac{\varphi-\varphi_d}{z^2-1}\right]
\right\}
\exp \left(\varphi\right).
\label{eq:ne-oml}\\
\frac{n_i(z)}{n_{i0}}
= &
\sqrt{-\frac{Z\beta\varphi}{\pi}}
  \left[1 + \sqrt{1 - \frac{\varphi_d}{z^2\varphi}}\right] 
+ \frac{e^{-Z\beta\varphi}}{2} 
\left[
1 - {\rm Erf}\left(\sqrt{-Z\beta\varphi}\right)\right]
 + \frac{\sqrt{1- z^{-2}}}{2}
e^{- Z\beta\tilde{\varphi}} 
\left[ 1 - {\rm Erf}\left(\sqrt{-Z\beta\tilde{\varphi}}\right) \right],
  \label{eq:ni-oml-alpert}
\end{align}
\end{widetext}
where 
\begin{align}
& z\equiv r/r_d, \\
& \varphi\equiv e\phi/k_B T_e, \\
& \beta\equiv
T_e/T_i, \\
& \tilde{\varphi} \equiv \left(\varphi -
\varphi_d/z^2\right)/(1-z^{-2}).
\end{align} 
The failure of OML theory manifests in an imaginary ion density when
$\phi > \phi_d/z^2.$ For the initially faster than $1/r^2$ increase in
$\phi(r)$ due to Debye shielding, which is the cause for ion
absorption radius at $r_m>r_d$ as noted previously for a negatively
charged dust, Allen et al. concluded that the contradiction between
$\phi(r)$ and an imaginary $n_i$ in Eq.~(\ref{eq:ni-oml-alpert}) would
be an intrinsic defect which prevents the OML theory from predicting
the plasma response.  Since the dust charge is related to the normal
electric field at the dust surface, which requires the solution of
$\phi(r),$ one reaches the inevitable position from
Ref.~\cite{allen-etal-jpp-2000} that OML theory can not predict the
dust charge, despite its success on dust potential as re-affirmed in
Ref.~\cite{lampe-jpp-2001,willis-etal-psst-2010}.

\section{Resolving the OML contradiction between $\phi(r)$ and $n_i(r)$\label{sec:corrected-oml}}

Intuitively it is quite puzzling that the OML simplification of
ignoring the absorption radius effect, which contributes a small error
in the ion charging current, would produce an imaginary ion density,
as revealed in Allen et al.'s analysis.  Physically, a maxima in ion
$\Phi_{\rm eff}(r)$ at $r_m>r_d$ would pose a barrier that turn back
ions with a range of $J.$ Ignoring this effect with the OML
approximation should manifest in an over-estimation of the ion density
for $r<r_m.$ So why Allen et al. discovered an apparently inherent
contradiction in the OML theory between $\phi(r)$ and $n_i^{OML}(r)?$
  
Following the physical picture just given, one is tempted to conclude
that the resolution has to come from a revision of the OML expression
for $n_i^{OML}(r)$ as given in Eq.~(\ref{eq:ni-oml-alpert}).  We find
that this is indeed the case.  The cause is a change in integration
bound for the OML ion density when the plasma potential transitions
from $\phi(r) < \phi_d/z^2$ to $\phi(r) > \phi_d/z^2.$ To understand
this subtlety, which has been evidently elusive for the past five
decades, we recall that since OML assumes a monotonically varying
$\Phi_{\rm eff}(r),$ the ions at $r$ have a collisionless orbit that
will either intercept the dust particle or be reflected by the
effective potential before it can reach the dust surface.

\begin{figure*}
\vspace{0.25cm}
\begin{centering}
\includegraphics[scale=0.85]{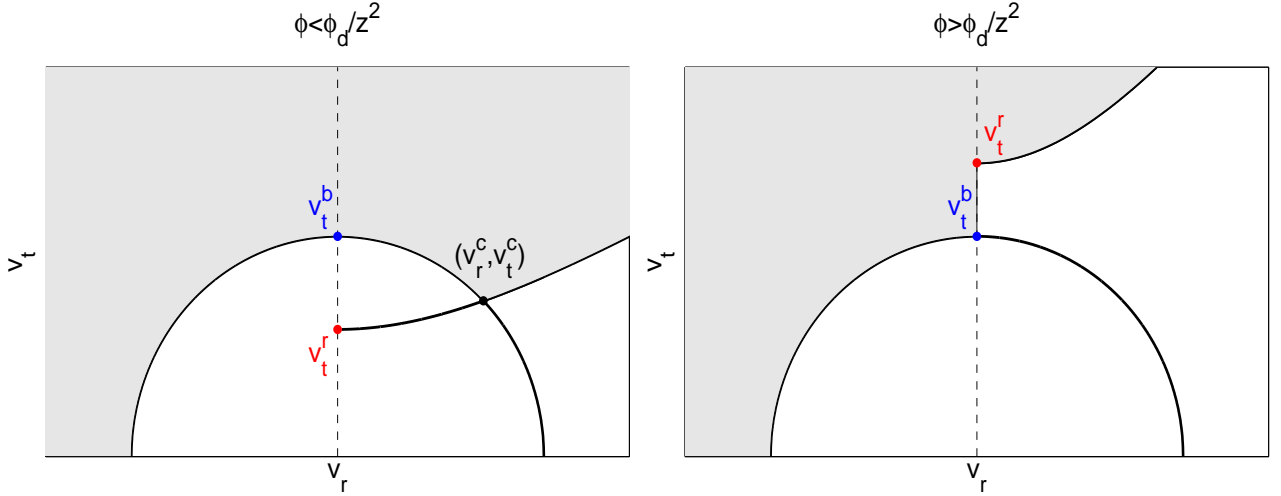}
\par\end{centering}
\caption{The integration bound in velocity space for the OML ion
  density, $n_i(z),$ changes depending on whether $\phi(z) <
  \phi_d/z^2$ or not. This results in two distinct forms for the ion density in OML theory.}
\label{fig:oml-integration-bounds}
\end{figure*}

In the canonical case that $\lim_{r\rightarrow\infty}\phi(r) = 0,$
the birth energy of the background plasma ion far away must have
\begin{align}
E_0=\frac{1}{2}m_i\left(v_r'^2 + v_t'^2\right) + Ze\phi(r\rightarrow\infty) \ge 0.
\end{align}
If this ion reaches $r,$ it must have, at $r,$ that
\begin{align}
E = \frac{1}{2}m_i\left(v_r^2 + v_t^2\right) + Ze\phi(r) = E_0 \ge 0.
\end{align}
For a negatively charged dust which has $\phi(r)< 0,$ a plasma ion of
such unbounded orbit (meaning that the ion orbit connects to infinity)
must have higher kinetic energy as it approaches the dust,
\begin{align}
v_r^2 + v_t^2 \ge - {2Ze}\phi(r)/m_i > 0.\label{eq:unbounded-orbit}
\end{align} 
As illustrated in Fig.~\ref{fig:oml-integration-bounds}, this is
outside a circle in $(v_r,v_t)$ space, which intercepts the $v_r=0$
axis at
\begin{align}
v_t^b = \sqrt{-{2Ze}\phi(r)/m_i}.\label{eq:vt-unbounded}
\end{align}

Not all of these ions can reach the dust surface ($r=r_d$) due to
angular momentum conservation.  In the OML approximation, the
effective potential $\Phi_{\rm eff}(r)$ peaks at $r=r_d,$
\begin{align}
\Phi_{\rm eff}(r_d) = {J^2}/{(2m_ir_d^2)} + Ze\phi_d. 
\end{align} 
The ions with
$E<\Phi_{\rm eff}(r_d)$ will be reflected by the effective potential
before they can reach $r_d.$ At $r>r_d,$ these reflected ions satisfy,
after explicitly writing out $E<\Phi_{\rm eff}(r_d),$
\begin{align} 
v_r^2 - \left(z^2 - 1 \right) v_t^2 < {2Ze} \left(\phi_d - \phi\right)/m_i.
\end{align}
For a negatively charged dust with $\phi_d - \phi(r) \le 0,$ the
reflected ions are bounded by a parabola in $(v_r,v_t)$ space,
\begin{align}
\left(z^2 - 1 \right) v_t^2 - v_r^2 > {2Ze}\left(\phi - \phi_d\right)/m_i.
\end{align}
As illustrated in Fig.~\ref{fig:oml-integration-bounds}, it intercepts
the $v_r=0$ axis at
\begin{align}
v_t^r =  \sqrt{\frac{2Ze}{m_i}\frac{\phi-\phi_d}{z^2 - 1}}.
\end{align}
As we shall see, this should be compared with the intercept of Eq.~(\ref{eq:unbounded-orbit})
with $v_r=0$ axis, i.e. $v_t^b$ in Eq.~(\ref{eq:vt-unbounded}).

To evaluate the ion density at $r\ge r_d,$ which is the solution of
the Vlasov equation, we integrate the Maxwellian distribution, which
is assumed for the background plasma far away, in the velocity space
region where they are allowed~\cite{alpert-book-1965}.  For ions with
$v_r<0,$ their population lies in the region of the $(v_r, v_t)$
velocity space given by
\begin{align}
v_r < 0 \,\,\, \& \,\,\,
v_r^2 + v_t^2 \ge - {2Ze} \phi(r)/m_i.
\end{align}
For ions with $v_r>0,$ only the reflected ions are present so
they are in a region of the velocity space given by
\begin{align}
& v_r \ge 0 \,\,\, \&\,\,\,
v_r^2 + v_t^2 \ge -{2Ze}\phi(r)/m_i \\
&
\left(z^2 - 1\right) v_t^2
-
v_r^2 > {2Ze}\left(\phi - \phi_d\right)/m_i
\end{align} 
These give rise to integration bounds for $v_r<0$ and $v_r>0$
separately.  The one for $v_r<0$ is straightforward, see
Fig.~\ref{fig:oml-integration-bounds}, and the previous OML
form~\cite{alpert-book-1965} is correct.

The one for $v_r>0$ is complicated by the possibility that
${v_t^r}^2 < {v_t^b}^2,$ which implies
\begin{align}
\phi < \phi_d/z^2.
\end{align}
If this is the case, the integration has two zones, separated by
a set of critical $v_r^c$ and $v_t^c,$ which are the intercept of the
two constraints,
$v_r^2 + v_t^2 = -{2Ze}\phi(r)/m_i$
and $\left(z^2 - 1\right) v_t^2
-
v_r^2 = {2Ze}\left(\phi - \phi_d\right)/m_i,$ 
\begin{align}
v_r^c = \sqrt{- \frac{2Ze}{m_i}\left(\phi - \frac{\phi_d}{z^2}\right)};\,\,\,
v_t^c = \sqrt{- \frac{2Ze}{m_i}\frac{\phi_d}{z^2}}
\end{align}
As shown in the left-side diagram of Fig.~\ref{fig:oml-integration-bounds}, the integration can
be carried out in two zones, depending on whether $v_r > v_r^c$ or
not.  The lower zone (denoted as I) is
\begin{align}
 v_r \in [0, v_r^c] \,\,\, \& \,\,\,
v_t \in [\sqrt{-({2Ze}/{m_i})\phi - v_r^2}, \infty).
\end{align}
The upper zone (denoted as II) is
\begin{align}
v_r \in [v_r^c, \infty) \,\,\, \& \,\,\,
v_t \in \left[\sqrt{ \frac{v_r^2 + \frac{2Ze}{m_i}\left(\phi -\phi_d\right)}{z^2 - 1}}, \infty\right).
\end{align}
The ion density is given by
\begin{align*}
n_i(r) =
 \frac{n_{i0}}{\sqrt{2\pi}} \left(\frac{m_i}{k_B T_i}\right)^{3/2}
 \iint_{{\rm I}+{\rm II}}
  e^{ -\frac{m_i\left(v_r^2 + v_t^2\right)+2Ze\phi}{2k_B T_i}
  } v_t dv_t dv_r,
\end{align*} 
which yields Eq.~(\ref{eq:ni-oml-alpert}), {\em now valid only for $\phi(r)<\phi_d/z^2$}.

\begin{figure}
\vspace{0.25cm}
\begin{centering}
\includegraphics[scale=0.3]{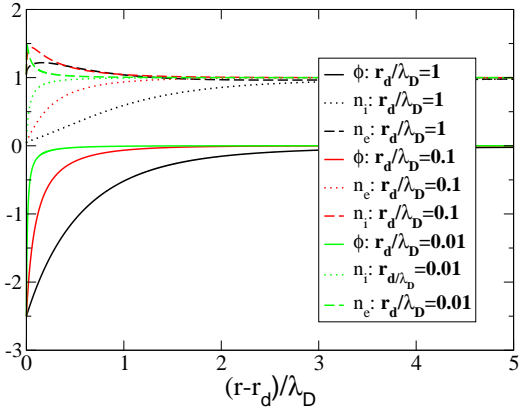}
\par\end{centering}
\caption{The plasma potential $\phi,$ ion density $n_i,$ and electron density $n_e$
are plotted as functions of $(r-r_d)/\lambda_D$ for three cases of
$r_d/\lambda_D=1$ (black), $r_d/\lambda_D=0.1$ (red), and $r_d/\lambda_D=0.01$ (green).}
\label{fig:oml-plasma-profile}
\end{figure}

In the case that ${v_t^r}^2 > {v_t^b}^2,$ i.e.,
\begin{align}
\phi > \phi_d/z^2,
\end{align}
the integration is restricted to a single zone given by
\begin{align}
v_r \in [0, \infty) \,\,\, \& \,\,\,
v_t \in \left[\sqrt{ \frac{v_r^2 + \frac{2Ze}{m_i}\left(\phi -\phi_d\right)}{z^2 - 1}}, \infty\right), \label{eq:v-bound-single}
\end{align}
which is illustrated in the right-side diagram of Fig.~\ref{fig:oml-integration-bounds}.
The ion density then takes the form
\begin{widetext}
\begin{align}
\frac{n_i(z)}{n_{i0}}
=
 \sqrt{-\frac{Z\beta\varphi}{\pi}}
+ \frac{e^{- Z\beta\varphi}}{2}
\left[
1 - {\rm Erf}\left(\sqrt{- Z\beta\varphi}\right)\right]
+ \frac{\sqrt{1- z^{-2}}}{2}
  e^{- Z\beta\tilde{\varphi}},\,\,\,\,\,\, \phi\ge\phi_d/z^2,
      \label{eq:ni-oml-corrected}
\end{align}
\end{widetext}
where $\phi(z)\ge \phi_d/z^2,$ while Eq.~(\ref{eq:ni-oml-alpert})
should be used for $\phi(z) < \phi_d/z^2.$ Contrasting
Eq.~(\ref{eq:ni-oml-corrected}) with Eq.~(\ref{eq:ni-oml-alpert}), one
sees that the previously-known imaginary OML ion density for
$\phi>\phi_d/z^2$ is removed in the corrected OML theory for $n_i.$

\section{OML prediction of plasma response and dust charge\label{sec:oml-result}} 

With the corrected OML ion density, the OML Poisson
equation~(\ref{eq:oml-poisson}) can be solved for the plasma response.
It is important to note that the radial electric field at the dust
surface $E_d=-d\varphi/d z(z=1)$ is related to the dust charge
according to Gauss's law, 
\begin{align}
Q_d = 4\pi \varepsilon_0 r_d^2 E_d.
\end{align}

Since the OML theory to date can not predict $Q_d,$ users of the OML charging theory
have been using a simple relation for spherical dust capacitance
due to Whipple~\cite{whipple-rpp-1981},
\begin{align}
Q_d = 4\pi\varepsilon_0 {r_d}\left(1 + {r_d}/{\lambda_D}\right)\phi_d,\label{eq:wipple-formula}
\end{align}
with 
\begin{align}
\lambda_D\equiv\sqrt{\varepsilon_0 T_e/e^2/n_{e0}}
\end{align}
the electron Debye length.
Whipple's idea is to expand the plasma density at $z\gg 1,$ as in the standard Debye shielding
calculation.
Applying this approach to OML's Poisson equation, one finds
\begin{align}
& n_e/n_{e0}\approx 1 + \varphi, \\
& n_i/n_{i0}\approx 1 - Z\beta\varphi,
\end{align}
and hence
\begin{align}
\frac{1}{z^2}\frac{d}{d z} \left(z^2\frac{d\varphi}{d z}\right)
=
\frac{r_d^2}{\lambda_D^2} \left(1 + Z\beta\right)\varphi.
\end{align}
Introducing the so-called linearized Debye length
\begin{align}
\lambda_{lin} \equiv {\lambda_D}/{\sqrt{1 + Z \beta}},
\end{align}
and defining the normalized dust radius as
\begin{align}
\hat{r_d} \equiv {r_d}/{\lambda_{lin}},
\end{align}
one finds the solution to the Poisson equation,
\begin{align}
\varphi = (\varphi_d/z)\exp\left[-\hat{r_d}(z-1)\right].\label{eq:oml-debye-shielding}
\end{align}
Hence the dust charge is
\begin{align}
  Q_d = 4\pi\varepsilon_0 {r_d} \left(1 +
    \frac{r_d\sqrt{1+ZT_e/T_i}}{\lambda_D}\right)\phi_d.
\label{eq:oml-wipple-formula}
\end{align}
Unlike the original Whipple formula [Eq.~(\ref{eq:wipple-formula})]
which assumes a uniform ion density, Eq.~(\ref{eq:oml-wipple-formula})
takes into account the ion density response.  It is interesting to
note that for a small dust $r_d \ll \lambda_D,$ the Debye shielding
contribution, which is usually small, can be enhanced substantially if
$Z\gg 1$ and/or $T_e\gg T_i.$ We will call the charge-potential
relationship given in Eq.~(\ref{eq:oml-wipple-formula}) the
generalized Whipple approximation, to distinguish it from the well-known
Whipple formula in Eq.~(\ref{eq:wipple-formula}).

\begin{figure}
\begin{centering}
\includegraphics[scale=0.3]{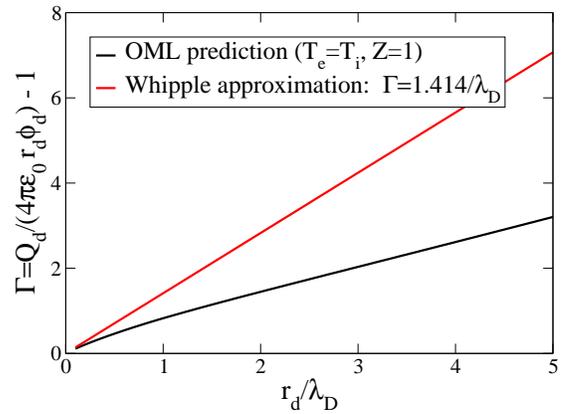}
\par\end{centering}
\caption{Dust capacitance is shown as a function of $r_d/\lambda_D.$
The deviation from Eq. (\ref{eq:oml-wipple-formula}) becomes large when the dust size becomes
comparable or greater than the Debye length.}
\label{fig:oml-scan-dust-size}
\end{figure}

The actual OML density for $r-r_d < \lambda_D$ can deviate
significantly from the asymptotic expansion valid when $r-r_d \gg
\lambda_D,$ see Fig.~\ref{fig:oml-plasma-profile}.  This is already
evident from the analytical form of $n_e$ and $n_i$ in OML theory.
Since the dust charge can be alternatively computed by integrating the
net charge of the plasma,
\begin{align}
Q_d = - \int (Zn_i - n_e)/{n_{e0}}d^3{\x},
\end{align}
due to overall charge conservation, one suspects that the OML
prediction of dust charge can be significantly different from that in
Eq.~(\ref{eq:oml-wipple-formula}).  From the numerically evaluated
$n_e$ and $n_i$ shown in Fig.~\ref{fig:oml-plasma-profile}, we find
that sharper deviation occurs closer to $r_d$ as the dust size becomes
smaller, but there is a greater spatial extent of the deviation for
large dust size.  Since the dust charge corresponds to the spatial
integration of electron and ion density, one finds that such deviation
from the simple Debye shielding calculation is proportional to
$r_d/\lambda_D.$ In Fig.~\ref{fig:oml-scan-dust-size}, we plot
\begin{align}
\Gamma\equiv Q_d/(4\pi \varepsilon_0 r_d \phi_d)-1
\end{align} 
as a function of
$r_d/\lambda_D.$ The specific case has $Z=1$ and $\beta=T_e/T_i=1,$ so
the generalized Whipple approximation, Eq.~(\ref{eq:oml-wipple-formula}), is simply
\begin{align}
\Gamma = 1.414 r_d/\lambda_D.
\end{align}
 The deviation of the OML prediction of
dust charge from Whipple approximation can be significantly greater if
the dust size is approaching the Debye shielding length, which is
consistent with the results of Ref. \cite{daugherty92}.  This is
ultimately due to the effect of angular momentum conservation on the
plasma density near the dust particle. Obviously a significant
correction in $Q_d$ implies a substantial change in the electrical
force $Q_d {\bf E},$ which can impact the dust dynamics, for example,
in a tokamak
reactor~\cite{tang-delzanno-jfe-2007,delzanno-tang-pop-2014}.

\section{summary\label{sec:summary}}

In conclusion, we have resolved a long-standing issue in the OML
charging theory, and by doing so, obtained a complete OML theory that
predicts both dust potential and dust charge. This is enabled by a
revised OML ion density formula for the case of $\phi>\phi_d/z^2,$
which is given in Eq.~(\ref{eq:ni-oml-corrected}). We also provide the
first calculation of the plasma potential and the dust charge using
the OML theory.  Our results show that for applications where the dust
particulate radius is much smaller than the plasma Debye length, the
Whipple approximation is in good agreement with the OML prediction of
the dust charge-potential relationship.  When the dust size is
comparable to or larger than the Debye shielding length, which is a
case of importance to magnetic fusion, we discover significant
deviation from the Whipple approximation of the dust charge-potential
relationship. This is attributed to the fundamental role of angular
momentum conservation in setting the plasma electron and ion density
near the dust particle.

\textcolor{black}{Since OML is an approximation of the OM theory by
  ignoring the ion absorption radius effect, one expects that its
  prediction of dust potential and charge should deviate from that of
  the OM theory. The usefulness of OML charging theory traditionally
  lies with its surprisingly good agreement with OM on dust
  potential~\cite{lampe-jpp-2001,kennedy03,delzanno04,willis-etal-psst-2010,delzanno-etal-ieee-2013}.
  With the new OML formulation that is able to calculate dust charge
  and plasma reponse, it is an important next step to carry out a
  detailed comparison of the OML dust charge prediction with OM theory
  for a range of dust sizes. }

\acknowledgments

This research was supported by the U.S. Department of Energy Office of
Science, Office of Fusion Energy Sciences, under the auspices of the
National Nuclear Security Administration of the U.S. Department of
Energy by Los Alamos National Laboratory, operated by Los Alamos
National Security LLC under contract DE-AC52-06NA25396.


\begin{thebibliography}{27}%
\makeatletter
\providecommand \@ifxundefined [1]{%
 \@ifx{#1\undefined}
}%
\providecommand \@ifnum [1]{%
 \ifnum #1\expandafter \@firstoftwo
 \else \expandafter \@secondoftwo
 \fi
}%
\providecommand \@ifx [1]{%
 \ifx #1\expandafter \@firstoftwo
 \else \expandafter \@secondoftwo
 \fi
}%
\providecommand \natexlab [1]{#1}%
\providecommand \enquote  [1]{``#1''}%
\providecommand \bibnamefont  [1]{#1}%
\providecommand \bibfnamefont [1]{#1}%
\providecommand \citenamefont [1]{#1}%
\providecommand \href@noop [0]{\@secondoftwo}%
\providecommand \href [0]{\begingroup \@sanitize@url \@href}%
\providecommand \@href[1]{\@@startlink{#1}\@@href}%
\providecommand \@@href[1]{\endgroup#1\@@endlink}%
\providecommand \@sanitize@url [0]{\catcode `\\12\catcode `\$12\catcode
  `\&12\catcode `\#12\catcode `\^12\catcode `\_12\catcode `\%12\relax}%
\providecommand \@@startlink[1]{}%
\providecommand \@@endlink[0]{}%
\providecommand \url  [0]{\begingroup\@sanitize@url \@url }%
\providecommand \@url [1]{\endgroup\@href {#1}{\urlprefix }}%
\providecommand \urlprefix  [0]{URL }%
\providecommand \Eprint [0]{\href }%
\providecommand \doibase [0]{http://dx.doi.org/}%
\providecommand \selectlanguage [0]{\@gobble}%
\providecommand \bibinfo  [0]{\@secondoftwo}%
\providecommand \bibfield  [0]{\@secondoftwo}%
\providecommand \translation [1]{[#1]}%
\providecommand \BibitemOpen [0]{}%
\providecommand \bibitemStop [0]{}%
\providecommand \bibitemNoStop [0]{.\EOS\space}%
\providecommand \EOS [0]{\spacefactor3000\relax}%
\providecommand \BibitemShut  [1]{\csname bibitem#1\endcsname}%
\let\auto@bib@innerbib\@empty
\bibitem [{\citenamefont {Langmuir}(1928)}]{langmuir-pnas-1928}%
  \BibitemOpen
  \bibfield  {author} {\bibinfo {author} {\bibfnamefont {I.}~\bibnamefont
  {Langmuir}},\ }\href {\doibase 10.1073/pnas.14.8.627} {\bibfield  {journal}
  {\bibinfo  {journal} {Proceedings of the National Academy of Sciences of the
  United States of America}\ }\textbf {\bibinfo {volume} {14}},\ \bibinfo
  {pages} {627} (\bibinfo {year} {1928})}\BibitemShut {NoStop}%
\bibitem [{\citenamefont {Mott-Smth}\ and\ \citenamefont
  {Langmuir}(1926)}]{mott-smith-langmuir-pr-1926}%
  \BibitemOpen
  \bibfield  {author} {\bibinfo {author} {\bibfnamefont {H.}~\bibnamefont
  {Mott-Smth}}\ and\ \bibinfo {author} {\bibfnamefont {I.}~\bibnamefont
  {Langmuir}},\ }\href {\doibase 10.1103/PhysRev.28.727} {\bibfield  {journal}
  {\bibinfo  {journal} {Physical Review}\ }\textbf {\bibinfo {volume} {28}},\
  \bibinfo {pages} {0727} (\bibinfo {year} {1926})}\BibitemShut {NoStop}%
\bibitem [{\citenamefont {Hutchinson}(2005)}]{hutchinson-book-2005}%
  \BibitemOpen
  \bibfield  {author} {\bibinfo {author} {\bibfnamefont {I.~H.}\ \bibnamefont
  {Hutchinson}},\ }\href@noop {} {\emph {\bibinfo {title} {Principles of Plasma
  Diagnostics}}}\ (\bibinfo  {publisher} {Cambridge University Press},\
  \bibinfo {address} {Cambridge},\ \bibinfo {year} {2005})\BibitemShut
  {NoStop}%
\bibitem [{\citenamefont {Shukla}\ and\ \citenamefont
  {Mamun}(2001)}]{shukla-book-2001}%
  \BibitemOpen
  \bibfield  {author} {\bibinfo {author} {\bibfnamefont {P.~K.}\ \bibnamefont
  {Shukla}}\ and\ \bibinfo {author} {\bibfnamefont {A.~A.}\ \bibnamefont
  {Mamun}},\ }\href@noop {} {\emph {\bibinfo {title} {Introduction to Dusty
  Plasma Physics}}}\ (\bibinfo  {publisher} {Institute of Physics Publishing},\
  \bibinfo {address} {Philadelphia},\ \bibinfo {year} {2001})\BibitemShut
  {NoStop}%
\bibitem [{\citenamefont {Fortov}\ and\ \citenamefont
  {Morfill}(2010)}]{fortov-book-2010}%
  \BibitemOpen
  \bibfield  {author} {\bibinfo {author} {\bibfnamefont {V.~E.}\ \bibnamefont
  {Fortov}}\ and\ \bibinfo {author} {\bibfnamefont {G.~E.}\ \bibnamefont
  {Morfill}},\ }\href@noop {} {\emph {\bibinfo {title} {Complex and Dusty
  Plasmas: From Laboratory to Space}}}\ (\bibinfo  {publisher} {CRC Press},\
  \bibinfo {address} {Boca Raton},\ \bibinfo {year} {2010})\BibitemShut
  {NoStop}%
\bibitem [{\citenamefont {Morfill}\ and\ \citenamefont
  {Ivlev}(2009)}]{morfill-rmp-2009}%
  \BibitemOpen
  \bibfield  {author} {\bibinfo {author} {\bibfnamefont {G.~E.}\ \bibnamefont
  {Morfill}}\ and\ \bibinfo {author} {\bibfnamefont {A.~V.}\ \bibnamefont
  {Ivlev}},\ }\href {\doibase 10.1103/RevModPhys.81.1353} {\bibfield  {journal}
  {\bibinfo  {journal} {Reviews of Modern Physics}\ }\textbf {\bibinfo {volume}
  {81}},\ \bibinfo {pages} {1353} (\bibinfo {year} {2009})}\BibitemShut
  {NoStop}%
\bibitem [{\citenamefont {Whipple}(1981)}]{whipple-rpp-1981}%
  \BibitemOpen
  \bibfield  {author} {\bibinfo {author} {\bibfnamefont {E.}~\bibnamefont
  {Whipple}},\ }\href@noop {} {\bibfield  {journal} {\bibinfo  {journal}
  {Reports on progress in Physics}\ }\textbf {\bibinfo {volume} {44}},\
  \bibinfo {pages} {1197} (\bibinfo {year} {1981})}\BibitemShut {NoStop}%
\bibitem [{\citenamefont {Verheest}(2000)}]{verheest-book-2000}%
  \BibitemOpen
  \bibfield  {author} {\bibinfo {author} {\bibfnamefont {F.}~\bibnamefont
  {Verheest}},\ }\href@noop {} {\emph {\bibinfo {title} {Waves in Dusty Space
  Plasmas}}}\ (\bibinfo  {publisher} {Kluwer Academic Publishers},\ \bibinfo
  {address} {Dordrecht},\ \bibinfo {year} {2000})\BibitemShut {NoStop}%
\bibitem [{\citenamefont {Draine}(2003)}]{draine-araa-2003}%
  \BibitemOpen
  \bibfield  {author} {\bibinfo {author} {\bibfnamefont {B.}~\bibnamefont
  {Draine}},\ }\href {\doibase 10.1146/annurev.astro.41.011802.094840}
  {\bibfield  {journal} {\bibinfo  {journal} {Annual Review of Astronomy and
  Astrophysics}\ }\textbf {\bibinfo {volume} {41}},\ \bibinfo {pages} {241}
  (\bibinfo {year} {2003})}\BibitemShut {NoStop}%
\bibitem [{\citenamefont {Mendis}\ and\ \citenamefont
  {Rosenberg}(1994)}]{mendis94}%
  \BibitemOpen
  \bibfield  {author} {\bibinfo {author} {\bibfnamefont {D.~A.}\ \bibnamefont
  {Mendis}}\ and\ \bibinfo {author} {\bibfnamefont {M.}~\bibnamefont
  {Rosenberg}},\ }\href@noop {} {\bibfield  {journal} {\bibinfo  {journal}
  {Annual Review of Astronomy and Astrophysics}\ }\textbf {\bibinfo {volume}
  {32}},\ \bibinfo {pages} {419} (\bibinfo {year} {1994})}\BibitemShut
  {NoStop}%
\bibitem [{\citenamefont {Bouchoule}(1999)}]{bouchoule-book-1999}%
  \BibitemOpen
  \bibfield  {author} {\bibinfo {author} {\bibfnamefont {A.}~\bibnamefont
  {Bouchoule}},\ }\href@noop {} {\emph {\bibinfo {title} {Dusty Plasmas:
  Physics, Chemistry and Technological Impacts in Plasma Processing}}}\
  (\bibinfo  {publisher} {Wiley},\ \bibinfo {address} {Chichester},\ \bibinfo
  {year} {1999})\BibitemShut {NoStop}%
\bibitem [{\citenamefont {Bacharis}\ \emph {et~al.}(2010)\citenamefont
  {Bacharis}, \citenamefont {Coppins},\ and\ \citenamefont
  {Allen}}]{bacharis10}%
  \BibitemOpen
  \bibfield  {author} {\bibinfo {author} {\bibfnamefont {M.}~\bibnamefont
  {Bacharis}}, \bibinfo {author} {\bibfnamefont {M.}~\bibnamefont {Coppins}}, \
  and\ \bibinfo {author} {\bibfnamefont {J.~E.}\ \bibnamefont {Allen}},\
  }\href@noop {} {\bibfield  {journal} {\bibinfo  {journal} {Phys. Rev. E}\
  }\textbf {\bibinfo {volume} {82}},\ \bibinfo {pages} {026403} (\bibinfo
  {year} {2010})}\BibitemShut {NoStop}%
\bibitem [{\citenamefont {Krasheninnikov}\ \emph {et~al.}(2011)\citenamefont
  {Krasheninnikov}, \citenamefont {Smirnov},\ and\ \citenamefont
  {Rudakov}}]{krash11}%
  \BibitemOpen
  \bibfield  {author} {\bibinfo {author} {\bibfnamefont {S.~I.}\ \bibnamefont
  {Krasheninnikov}}, \bibinfo {author} {\bibfnamefont {R.~D.}\ \bibnamefont
  {Smirnov}}, \ and\ \bibinfo {author} {\bibfnamefont {D.~L.}\ \bibnamefont
  {Rudakov}},\ }\href@noop {} {\bibfield  {journal} {\bibinfo  {journal}
  {Plasma Physics and Controlled Fusion}\ }\textbf {\bibinfo {volume} {53}},\
  \bibinfo {pages} {083001} (\bibinfo {year} {2011})}\BibitemShut {NoStop}%
\bibitem [{\citenamefont {\relax{Ya}. L.~Al'pert}\ \emph
  {et~al.}(1965)\citenamefont {\relax{Ya}. L.~Al'pert}, \citenamefont
  {Gurevich},\ and\ \citenamefont {Pitaevskii}}]{alpert-book-1965}%
  \BibitemOpen
  \bibfield  {author} {\bibinfo {author} {\bibnamefont {\relax{Ya}.
  L.~Al'pert}}, \bibinfo {author} {\bibfnamefont {A.~V.}\ \bibnamefont
  {Gurevich}}, \ and\ \bibinfo {author} {\bibfnamefont {L.~P.}\ \bibnamefont
  {Pitaevskii}},\ }\href@noop {} {\emph {\bibinfo {title} {Space Physics with
  Artificial Satellites}}}\ (\bibinfo  {publisher} {Plenum Press, New York},\
  \bibinfo {year} {1965})\BibitemShut {NoStop}%
\bibitem [{\citenamefont {Laframboise}(1966)}]{laframboise-thesis-1966}%
  \BibitemOpen
  \bibfield  {author} {\bibinfo {author} {\bibfnamefont {J.}~\bibnamefont
  {Laframboise}},\ }\href@noop {} {\emph {\bibinfo {title} {Theory of spherical
  and cylindrical Langmuir probes in a collisionless, Maxwellian plasma at
  rest}}},\ \bibinfo {type} {Tech. Rep.}\ (\bibinfo  {institution} {Toronto
  Univ. (Ontario). Inst. for Aerospace Studies},\ \bibinfo {year}
  {1966})\BibitemShut {NoStop}%
\bibitem [{\citenamefont {Lampe}(2001)}]{lampe-jpp-2001}%
  \BibitemOpen
  \bibfield  {author} {\bibinfo {author} {\bibfnamefont {M.}~\bibnamefont
  {Lampe}},\ }\href
  {http://permalink.lanl.gov/object/view?what=info:lanl-repo/isi/000171401000002}
  {\bibfield  {journal} {\bibinfo  {journal} {Journal of Plasma Physics}\
  }\textbf {\bibinfo {volume} {65}},\ \bibinfo {pages} {171} (\bibinfo {year}
  {2001})}\BibitemShut {NoStop}%
\bibitem [{\citenamefont {Delzanno}\ \emph {et~al.}(2004)\citenamefont
  {Delzanno}, \citenamefont {Lapenta},\ and\ \citenamefont
  {Rosenberg}}]{delzanno04}%
  \BibitemOpen
  \bibfield  {author} {\bibinfo {author} {\bibfnamefont {G.~L.}\ \bibnamefont
  {Delzanno}}, \bibinfo {author} {\bibfnamefont {G.}~\bibnamefont {Lapenta}}, \
  and\ \bibinfo {author} {\bibfnamefont {M.}~\bibnamefont {Rosenberg}},\
  }\href@noop {} {\bibfield  {journal} {\bibinfo  {journal} {Physical Review
  Letters}\ }\textbf {\bibinfo {volume} {92}},\ \bibinfo {pages} {350021}
  (\bibinfo {year} {2004})}\BibitemShut {NoStop}%
\bibitem [{\citenamefont {Willis}\ \emph {et~al.}(2010)\citenamefont {Willis},
  \citenamefont {Coppins}, \citenamefont {Bacharis},\ and\ \citenamefont
  {Allen}}]{willis-etal-psst-2010}%
  \BibitemOpen
  \bibfield  {author} {\bibinfo {author} {\bibfnamefont {C.~T.~N.}\
  \bibnamefont {Willis}}, \bibinfo {author} {\bibfnamefont {M.}~\bibnamefont
  {Coppins}}, \bibinfo {author} {\bibfnamefont {M.}~\bibnamefont {Bacharis}}, \
  and\ \bibinfo {author} {\bibfnamefont {J.~E.}\ \bibnamefont {Allen}},\
  }\href@noop {} {\bibfield  {journal} {\bibinfo  {journal} {Plasma Sources
  Science and Technology}\ }\textbf {\bibinfo {volume} {19}},\ \bibinfo {pages}
  {065022} (\bibinfo {year} {2010})}\BibitemShut {NoStop}%
\bibitem [{\citenamefont {Delzanno}\ \emph {et~al.}(2013)\citenamefont
  {Delzanno}, \citenamefont {Camporeale}, \citenamefont {Moulton},
  \citenamefont {Borovsky}, \citenamefont {MacDonald},\ and\ \citenamefont
  {Thomsen}}]{delzanno-etal-ieee-2013}%
  \BibitemOpen
  \bibfield  {author} {\bibinfo {author} {\bibfnamefont {G.~L.}\ \bibnamefont
  {Delzanno}}, \bibinfo {author} {\bibfnamefont {E.}~\bibnamefont
  {Camporeale}}, \bibinfo {author} {\bibfnamefont {J.~D.}\ \bibnamefont
  {Moulton}}, \bibinfo {author} {\bibfnamefont {J.~E.}\ \bibnamefont
  {Borovsky}}, \bibinfo {author} {\bibfnamefont {E.~A.}\ \bibnamefont
  {MacDonald}}, \ and\ \bibinfo {author} {\bibfnamefont {M.}~\bibnamefont
  {Thomsen}},\ }\href@noop {} {\bibfield  {journal} {\bibinfo  {journal} {IEEE
  Transactions on Plasma Science}\ }\textbf {\bibinfo {volume} {41}},\ \bibinfo
  {pages} {3577} (\bibinfo {year} {2013})}\BibitemShut {NoStop}%
\bibitem [{\citenamefont {Bohm}(1949)}]{bohm-etal-book-1949}%
  \BibitemOpen
  \bibfield  {author} {\bibinfo {author} {\bibfnamefont {D.}~\bibnamefont
  {Bohm}},\ }\href@noop {} {\emph {\bibinfo {title} {The Characteristics of
  Electrical Discharges in Magnetic Fields}}}\ (\bibinfo  {publisher}
  {Mc-Graw-Hill},\ \bibinfo {address} {New York},\ \bibinfo {year}
  {1949})\BibitemShut {NoStop}%
\bibitem [{\citenamefont {Allen}(1992)}]{allen-ps-1992}%
  \BibitemOpen
  \bibfield  {author} {\bibinfo {author} {\bibfnamefont {J.}~\bibnamefont
  {Allen}},\ }\href {\doibase 10.1088/0031-8949/45/5/013} {\bibfield  {journal}
  {\bibinfo  {journal} {Physica Scripta}\ }\textbf {\bibinfo {volume} {45}},\
  \bibinfo {pages} {497} (\bibinfo {year} {1992})}\BibitemShut {NoStop}%
\bibitem [{\citenamefont {Allen}\ \emph {et~al.}(2000)\citenamefont {Allen},
  \citenamefont {Annaratone},\ and\ \citenamefont
  {de~Angelis}}]{allen-etal-jpp-2000}%
  \BibitemOpen
  \bibfield  {author} {\bibinfo {author} {\bibfnamefont {J.}~\bibnamefont
  {Allen}}, \bibinfo {author} {\bibfnamefont {B.}~\bibnamefont {Annaratone}}, \
  and\ \bibinfo {author} {\bibfnamefont {U.}~\bibnamefont {de~Angelis}},\
  }\href@noop {} {\bibfield  {journal} {\bibinfo  {journal} {Journal of Plasma
  Physics}\ }\textbf {\bibinfo {volume} {63}},\ \bibinfo {pages} {299}
  (\bibinfo {year} {2000})}\BibitemShut {NoStop}%
\bibitem [{\citenamefont {Tang}\ and\ \citenamefont
  {Delzanno}(2010)}]{tang-delzanno-jfe-2007}%
  \BibitemOpen
  \bibfield  {author} {\bibinfo {author} {\bibfnamefont {X.~Z.}\ \bibnamefont
  {Tang}}\ and\ \bibinfo {author} {\bibfnamefont {G.~L.}\ \bibnamefont
  {Delzanno}},\ }\href {\doibase 10.1007/s10894-010-9295-x} {\bibfield
  {journal} {\bibinfo  {journal} {Journal of Fusion Energy}\ }\textbf {\bibinfo
  {volume} {29}},\ \bibinfo {pages} {407} (\bibinfo {year} {2010})}\BibitemShut
  {NoStop}%
\bibitem [{\citenamefont {Delzanno}\ and\ \citenamefont
  {Tang}(2014{\natexlab{a}})}]{delzanno-tang-pop-2014}%
  \BibitemOpen
  \bibfield  {author} {\bibinfo {author} {\bibfnamefont {G.~L.}\ \bibnamefont
  {Delzanno}}\ and\ \bibinfo {author} {\bibfnamefont {X.~Z.}\ \bibnamefont
  {Tang}},\ }\href@noop {} {\bibfield  {journal} {\bibinfo  {journal} {Physics
  of Plasmas}\ }\textbf {\bibinfo {volume} {21}},\ \bibinfo {eid} {022502}
  (\bibinfo {year} {2014}{\natexlab{a}})}\BibitemShut {NoStop}%
\bibitem [{\citenamefont {Delzanno}\ and\ \citenamefont
  {Tang}(2014{\natexlab{b}})}]{delzanno-tang-submitted-2014}%
  \BibitemOpen
  \bibfield  {author} {\bibinfo {author} {\bibfnamefont {G.~L.}\ \bibnamefont
  {Delzanno}}\ and\ \bibinfo {author} {\bibfnamefont {X.~Z.}\ \bibnamefont
  {Tang}},\ }\href@noop {} {\bibfield  {journal} {\bibinfo  {journal} {Phys.
  Rev. Lett.}\ }\textbf {\bibinfo {volume} {113}},\ \bibinfo {pages} {035002} (\bibinfo {year}
  {2014}{\natexlab{b}})}\BibitemShut {NoStop}%
\bibitem [{\citenamefont {Kennedy}\ and\ \citenamefont
  {Allen}(2003)}]{kennedy03}%
  \BibitemOpen
  \bibfield  {author} {\bibinfo {author} {\bibfnamefont {R.}~\bibnamefont
  {Kennedy}}\ and\ \bibinfo {author} {\bibfnamefont {J.}~\bibnamefont
  {Allen}},\ }\href {\doibase 10.1017/S0022377803002265} {\bibfield  {journal}
  {\bibinfo  {journal} {Journal of Plasma Physics}\ }\textbf {\bibinfo {volume}
  {69}},\ \bibinfo {pages} {485} (\bibinfo {year} {2003})}\BibitemShut
  {NoStop}%
\bibitem [{\citenamefont {Daugherty}\ \emph {et~al.}(1992)\citenamefont
  {Daugherty}, \citenamefont {Porteous}, \citenamefont {Kilgore},\ and\
  \citenamefont {Graves}}]{daugherty92}%
  \BibitemOpen
  \bibfield  {author} {\bibinfo {author} {\bibfnamefont {J.}~\bibnamefont
  {Daugherty}}, \bibinfo {author} {\bibfnamefont {R.}~\bibnamefont {Porteous}},
  \bibinfo {author} {\bibfnamefont {M.~D.}\ \bibnamefont {Kilgore}}, \ and\
  \bibinfo {author} {\bibfnamefont {D.}~\bibnamefont {Graves}},\ }\href@noop {}
  {\bibfield  {journal} {\bibinfo  {journal} {Journal of Applied Physics}\
  }\textbf {\bibinfo {volume} {72}},\ \bibinfo {pages} {3934} (\bibinfo {year}
  {1992})}\BibitemShut {NoStop}%
\end{thebibliography}

%

\appendix

\section{The breakdown of previous OML forumlation}

As pointed out by Allen et al~\cite{allen-etal-jpp-2000}, the OML ion
density can become imaginary if one is to solve the Poisson equation,
Eq.~(\ref{eq:oml-poisson}), for the plasma potential using the
previously known OML ion density formula by Al'pert et
al~\cite{alpert-book-1965}. This suggests a fundamental breakdown of
the OML theory for plasma potential and hence dust charge calculation.
Allen et al investigated this problem from the angle of the OML
approximation, i.e., the negelect of the ion absorption radius effect.
In the main text, we give a physics argument that ignoring the ion
absorption radius effect should only introduce a discrepancy in ion
density compared to OM prediction, but not an unphysical imaginary
number. Our corrected calculation of the OML ion density removes the
possibility of an imaginary density. Here we give a more detailed
account of how the imaginary ion density comes about in the previous
OML formulation.  The objective is to further clarify the contrast
between (1) a physical approximation (OML) that introduces
quantitative discrepency which vanishes in its limit of applicability,
and (2) an invalid theoretical formulation that produces unphysical
results.

The condition for $n_i(z)$ to become imaginary (hence
  unphysical) in Eq.~(\ref{eq:ni-oml-alpert}) is to have
\begin{align}
  \frac{\varphi_d}{z^2\varphi} > 1. \label{eq:cross-over}
\end{align}
For a negatively charged dust ($\varphi<0$ and $\varphi_d<0$), this
implies a plasma potential $\varphi(z)$ that rises faster than
$\varphi_d/z^2$ with $z-1$ the normalized distance from the dust
surface.  As long as Debye shielding is in action (i.e. the plasma
transport to the dust surface is not intrinsically ambipolar), we know
that $\varphi$ would rise exponentially as a function of $z$ within
the Debye shielding sphere.  The cross-over point after which
Eq.~(\ref{eq:cross-over}) is satisfied can be illustrated using the
Debye shielding potential solution of
Eq.~(\ref{eq:oml-debye-shielding}) by setting
$\varphi_d/z^2\varphi=1.$ The cross-over point $z_c$ is the solution
of
\begin{align}
z_c\exp\left[-\hat{r_d}(z_c -1)\right] = 1. \label{eq:zc}
\end{align}
Here $\hat{r_d}$ is the dust radius normalized against the Debye
length and $z-1$ is the radial distance from the dust surface
normalized against the dust size.  For $\hat{r_d}=0.01,$ one finds
$z_c=648$ or 6.48 Deybe length away.  If $\hat{r_d}=0.0001,$ we have
$z_c=116672$ or 11.7 Debye length away.  In other words, the breakdown
of the ion density formula, Eq.~(\ref{eq:ni-oml-alpert}), is the
direct result of the Debye shielding physics, which is independent of
how small the dust size is compared to the Debye length.

Of course, the Debye shielding potential as given in
Eq.~(\ref{eq:oml-debye-shielding}) is an analytical approximation, so
the actual cross-over point $z_c$ would likely differ in its exact
location from that predicted by Eq.~(\ref{eq:zc}).  When the OML
Poisson equation is numerically solved, the failure of the OML
formulation using the ion density given in
Eq.~(\ref{eq:ni-oml-alpert}) manifests in a not-a-number floating
point violation for all small $\hat{r_d}$s we have attempted, as one
would expect by taking the squared root of a negative number.  This is
to be contrasted with the corrected OML formulation which produces the
numerical solutions shown in section~\ref{sec:oml-result}.

\end{document}